\documentclass[journal]{IEEEtran}
\usepackage{amsmath,mathtools,amsfonts,amssymb,amsthm}
\usepackage{cite}
\usepackage{lipsum}
\usepackage{url}
\newtheorem{proposition}{Proposition}
\usepackage{dsfont}
\usepackage{booktabs}
\usepackage{enumitem}

\usepackage{color}
\usepackage{xcolor}

\definecolor{NoteColor}{RGB}{0,104,149}

\newcommand{\expected}[1]{{\leavevmode\mathbb{E}\left[#1\right]}}

\usepackage[compact]{titlesec}         
\titlespacing{\section}{0pt}{0pt}{0pt} 
\AtBeginDocument{
  \setlength\abovedisplayskip{0pt}
  \setlength\belowdisplayskip{0pt}}

\ifCLASSINFOpdf
\else
\fi

\begin{document}



\title{Physical and Economic Viability of Cryptocurrency Mining for Provision of Frequency Regulation: \\ A Real-World Texas Case Study}



\author{

    Rayan~El~Helou$^\ast$, Ali~Menati$^\ast$ and Le~Xie$^{\dagger,~\S}$ \\
    \{$^\ast$\IEEEmembership{Student Member}, $^\dagger$\IEEEmembership{Fellow}\},~IEEE

    \thanks{Rayan~El~Helou, Ali~Menati and Le~Xie are with the Department of Electrical and Computer Engineering, Texas A\&M University, College Station, TX, USA. Corresponding Author: Le Xie, email: le.xie@tamu.edu
    }

}

\maketitle



\begin{abstract}
Demand flexibility plays a pivotal role  in modern power systems with high penetration of variable energy resources. In recent years, one of the fastest-growing  flexible energy demands has been proof-of-work-based cryptocurrency mining facilities. Due to their competitive ramping capabilities and demonstrated flexibility, such fast-responding loads are capable of participating in frequency regulation services for the grid while simultaneously increasing their own operational revenue. In this paper, we investigate the physical and economic viability of employing cryptocurrency mining facilities to provide  frequency regulation in large power systems. We quantify mining facilities' operational profit, and propose a decision-making framework to explore their optimal participation strategy and account for the most influential factors. We employ real-world ERCOT ancillary services data in our case study to investigate the conditions under which provision of frequency regulation in the Texas grid is profitable. We also perform transient level simulations using a synthetic Texas grid to demonstrate the competitiveness of mining facilities at frequency regulation provision.
\end{abstract}


\IEEEpeerreviewmaketitle


\section{Introduction}
\IEEEPARstart{O}{ne} of the primary roles of an electric transmission grid operator is to simultaneously balance supply and demand and ensure reliable grid operations. This has become a challenging task in recent years with the proliferation of distributed energy resources, such as renewables and electric vehicles \cite{wang2019chance}, which introduce increasing levels of uncertainty to grid operations. Inevitably, ancillary services are bound to become a cornerstone to the resilience of modern power systems, provided that they can offer fast and affordable flexibility. Among the most pivotal ancillary services is frequency regulation, which is a fast time-scale control service that accounts for electric supply-demand imbalance not covered by the slower time-scale energy market \cite{nitika}. Large frequency deviations may harm grid infrastructure, trip under-frequency relays, and lead to power system failure \cite{WU2017428, 8327538, BUSBY2021102106}.

Traditionally, frequency regulation is done through generation facilities capable of quickly responding either to local frequency measurements or to signals communicated by the grid operator to increase and decrease their power output. In recent years, high elasticity of flexible loads has been utilized to help balance supply and demand and provide frequency regulation in the grid. Among such flexible loads is the emerging cryptocurrency mining facilities which create a unique opportunity to offer frequency regulation and other ancillary services \cite{menati2023optimization}. In comparison, conventional methods for provision of frequency regulation, such as thermal fuel and hydro generators, are expensive and rather slow \cite{hao2015potentials}, but cryptocurrency mining facilities can demonstrably provide significantly faster response \cite{lancium}.

An alternative to cryptocurrency mining facilities is Internet data centers, which are computing-based loads that have also been used to participate in ancillary services. They are faster than conventional generators and are usually paired with battery storage systems in order to provide frequency regulation \cite{shi2017using, aksanli2014providing, li2014integrated}. The flexibility of internet data centers is harnessed by managing their delay-tolerant workload through proper hardware design and algorithmic solutions to adapt their processing power. There has been extensive research on data center load management techniques such as speed-scaling \cite{chen2015interaction}, power-capping \cite{capping13}, capacity right sizing \cite{sizing11}, and geographical load balancing \cite{6195508, 7218654}. However, such data centers are limited in their flexibility compared to cryptocurrency mining facilities, due to the former's time-sensitive computational obligations \cite{quirk2021cryptocurrency}. In practice, since cryptominers are not typically considered critical load, they are fully flexible and could potentially shut down all of their operational capacity within a few seconds \cite{menati2022modeling}. For example, during the winter storm Elliot in December 2022, cryptocurrency mining facilities in Texas provided 1,475 MW of load reduction, which was more than 96\% of their total capacity \cite{winterstorm}.

\begin{table}[htb]
    \setlength{\tabcolsep}{4pt}
    \centering
    \caption{Summary of 2022 Ancillary Services Market at ERCOT}
    \begin{tabular}{rcccc} \toprule
           \begin{tabular}[c]{@{}l@{}}Hourly Average \end{tabular} 
         & \begin{tabular}[c]{@{}l@{}}Reg-Up \end{tabular} 
         & \begin{tabular}[c]{@{}l@{}}Reg-Down\end{tabular}
         & \begin{tabular}[c]{@{}l@{}}RRS\end{tabular}
         & \begin{tabular}[c]{@{}l@{}}NSRS\end{tabular}
            \\ \midrule 

        Price for Capacity (\$/MW)           &   21.67 &      8.46 &      20.31 &      22.49 \\
        Capacity Procured (MW) &     359 &       348 &       2863 &       3895 \\
        Deployment Rate       &     16\% &       25\% &  $\approx$ 0\% &  $\approx$ 0\% \\
        
        \bottomrule
    \end{tabular}
    \label{tab:AS_statistics}
\end{table}

In the Electric Reliability Council of Texas (ERCOT), cryptocurrency mining loads are capable of performing all ancillary services and participating in real-time energy market \cite{drercot}. Some ancillary services like the Responsive Reserve Service (RRS) and the Non-spinning Reserve Service (NSRS) are designed for emergency reserve, and they would be deployed in instances of large generation shortages \cite{CLRercot}. Other ancillary services like Reg-Up/Down are used for frequency regulation, and as shown in Table \ref{tab:AS_statistics}, the size of Reg-Up/Down is much smaller than the bulk size of NSRS or RRS. In terms of deployment, it can also be seen that RRS and NSRS are rarely deployed, while Reg-Up/Down is deployed more often. The high flexibility and fast response time of cryptominers qualify them as a suitable candidate for frequency regulation provision, unlike other loads.

This paper investigates cryptominers' participation in frequency regulation from both economic and operational perspectives, and our main contributions are summarized as follows: 
\begin{enumerate}
    \item We present a decision-making framework for assessing the economic viability of cryptominers' participation in frequency regulation services considering the most influential factors including market clearing prices and deployment rates of these services, cryptocurrency prices, mining energy, and electricity prices. We utilize a historical record of real-world data to demonstrate under what circumstances participation is profitable in the Texas grid.
    
    \item We formulate the real-time frequency regulation dispatch problem from a grid operator's perspective subject to capacity and ramp constraints of resource entities. To solve the problem, we propose alternative real-time regulation dispatch mechanisms that address issues of fairness in allocation and efficiency of resource utilization when dispatching to multiple mining facilities at once.
    
    \item We perform transient-level simulations using a 2000-bus synthetic Texas grid to demonstrate the physical viability of cryptominers participation in frequency regulation services, with our results revealing substantial improvement in resiliency compared to existing alternatives.

\end{enumerate}

The remainder of the paper is organized as follows. In Section \ref{sec:market_formulation}, the frequency regulation market is formulated, where we present the parties involved, the day-ahead market, and real-time dispatch. In Section \ref{sec:profit_formulation}, profit by mining facilities is formulated, where we present revenue due to mining, profit due to frequency regulation, and a set of proposed strategies on when to participate in the frequency regulation market. In Section \ref{sec:case_study}, real-world data from ERCOT and historical records of Bitcoin prices are utilized to quantify and illustrate the feasibility and viability of participation by Bitcoin mining facilities. Finally, Section \ref{sec:conclusion} provides concluding remarks.

\vspace{1em}
\section{Regulation Market Formulation}
\label{sec:market_formulation}
In this section, we formulate the frequency regulation market, and we use ERCOT as an example for a possible set of rules and requirements adopted in the real-world. In particular, we introduce all the parties involved in this market and what role each plays, we formulate the day-ahead market (DAM) and the clearing process which yields hourly prices and capacities, and we formulate the real-time dispatch of all regulation service providers.

\begin{figure*}[!htbp]
\centering
\includegraphics[width=0.8\textwidth]{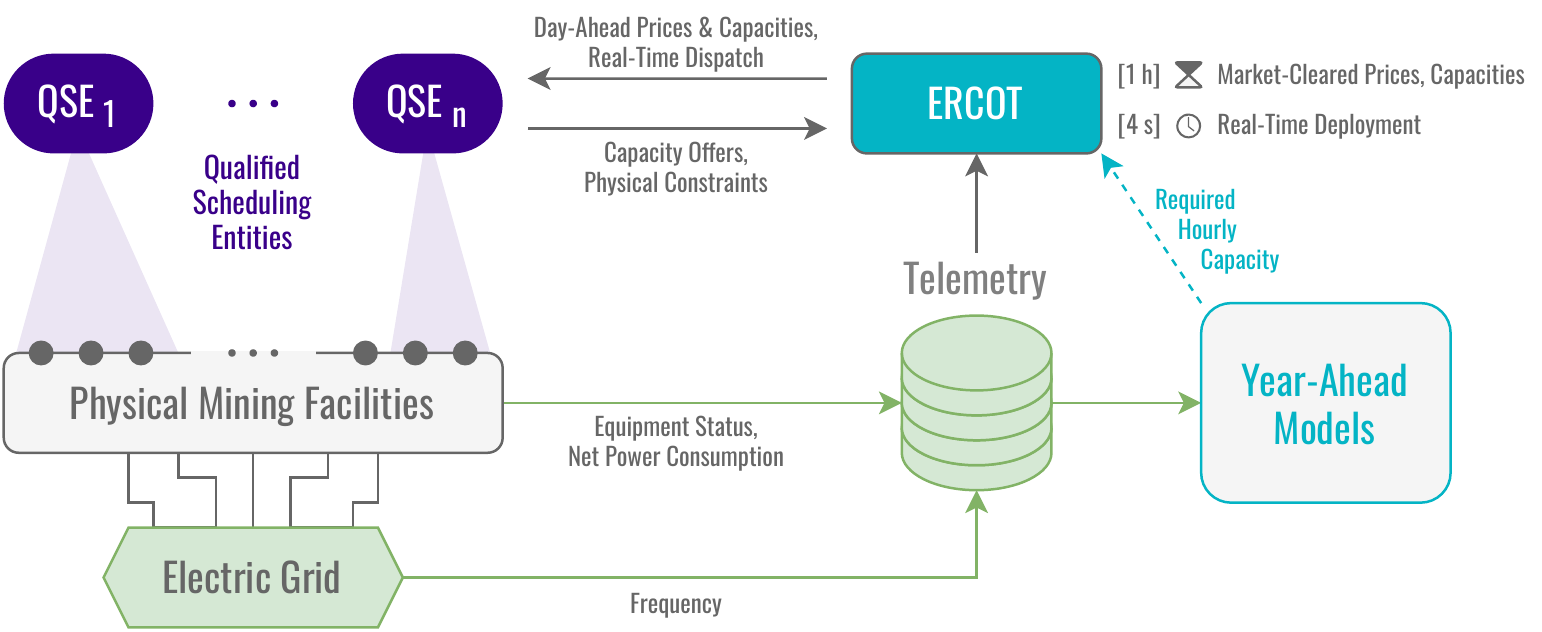}
\caption{Physical and economic interaction between ERCOT and the mining facilities participating in frequency regulation. Physical facilities interact with ERCOT via Qualified Scheduling Entities. In the Day-Ahead Market, prices and capacities are cleared hourly, while in Real-Time deployment, dispatch is updated every four seconds. Telemetry is used both for real-time control and for improving year-ahead models that determine future need for capacity.}
\label{fig:ercot_AS}
\end{figure*}

\subsection{Parties Involved}
There are four main parties involved in ancillary services markets: the market operator, the buyers, the sellers and the resource entities which provide the actual services, managed by the sellers. Fig. \ref{fig:ercot_AS} illustrates the interaction between these parties both in real-time and day-ahead.

Load Serving Entities (LSEs) provide electric service to customers and they serve as the buyers in this market, making them solely responsible for covering the costs of ancillary services. Even though LSEs pay for it, the regulation power demand is not explicitly determined by them, but by the market operator whose role is to ensure grid resiliency, which ultimately serves customers of the LSEs. Reference \cite{nitika} details methodologies for determining required hourly ancillary services demand in ERCOT. Each electric power transmission system in the United States is operated by a regional transmission organization subject to federal and local regulations. Within its region, such an organization serves as the market operator. In this paper, we focus on the case of Texas, where ERCOT is the market operator for the Texas interconnection. Finally, sellers, who manage the physical mining facilities considered in this paper, are responsible for submitting capacity offers in the day-ahead market, while ensuring that any cleared capacity is deliverable in real-time. In this paper, we use the ERCOT terminology Qualified Scheduling Entities (QSEs) to refer to such sellers who participate in the market on behalf of the mining facilities, as illustrated in Fig. \ref{fig:ercot_AS}.

In the day-ahead market, the market operator clears prices and capacities for ancillary services, and in real-time it observes frequency changes and dispatches suppliers of regulation services to increment or decrement power consumption or production in attempt to regulate frequency. Since demand is perfectly inelastic in ancillary service markets, the market operator needs to first determine exactly how much demand is needed for each period of time. In the case of ERCOT, demand for capacity at every hour of the year is determined prior to the beginning of each year and is fixed for one whole year \cite{nitika}. In Section \ref{sec:day_ahead}, we overview how prices and capacities for each resource entity are determined in the market clearing process, and in Section \ref{sec:real_time}, we overview how resources are dispatched to regulate frequency, and we propose alternative dispatch mechanisms that address fairness and resource utilization.

\subsection{Day-Ahead Market}
\label{sec:day_ahead}
In this subsection, we detail how prices and quantities for each seller are determined during the market clearing process. The term \textit{day-ahead} comes from the fact that regulation services are purchased as capacity in advance so that they can be deployed in the next day when needed. Here are some key features of the day-ahead market, based on ERCOT's protocols:
\begin{itemize}
    \item All sellers in the market are payed at the same hourly market clearing price (in \$/MW), i.e. no locational-prices.
    \item Hourly demand for capacity (in MW) is inelastic, determined year-ahead by forecasting renewables' uncertainty.
\end{itemize}
Frequency regulation is offered by multiple forms of ancillary services. In this paper, we focus only on two kinds of services, named in ERCOT as Reg-Up and Reg-Down. The former is deployed in real-time when the frequency is lower than desired and more generation (or less consumption) of power is needed, and vice versa for the latter. For each of those two, at the beginning of each day the market clearing process produces 24 prices (in \$/MW) for Reg-Up and 24 for Reg-Down, one for each hour of the upcoming day. Furthermore, each resource entity which submits an offer (as a seller) for each hour is told for that hour what its individual cleared capacity (in MW) is for both Reg-Up and Reg-Down.

The market clearing process is based on the principle that expected demand for capacity needs to be available at all times. In the year-ahead, as described in reference \cite{nitika}, ERCOT uses historic variations in loads and renewables to forecast frequency fluctuations in the upcoming year, which then translates to expected demand for capacity for each hour of the upcoming year, $\widehat{D^\text{up}}$ and $\widehat{D^\text{dn}}$, corresponding to Reg-Up and Reg-Down respectively. Using this inelastic demand, the market operator clears the market by considering the capacities and prices offered by all $n$ QSEs. Out of the $n$ offers submitted, $m\leq n$ are cleared for non-zero capacity. Let $c^\text{up}_i,c^\text{dn}_i ~ \forall i\in[1,m]$ be the cleared Reg-Up and Reg-Down capacities respectively. The market operator is expected to ensure that $\sum_{i=1}^m c^\text{up}_i = \widehat{D^\text{up}}$ and $\sum_{i=1}^m c^\text{dn}_i = \widehat{D^\text{dn}}$.

\subsection{Real-Time Dispatch}
\label{sec:real_time}
In this subsection, we detail how Reg-Up and Reg-Down power is dispatched for each resource entity during real-time operations, and we investigate alternative approaches for allocation across multiple entities. The term \textit{real-time} comes from the fact that regulation services are communicated and deployed in quick response to changes in frequency. Here are some key features of real-time operations pertaining to this study, based on ERCOT's protocols:
\begin{itemize}
    \item Regulation dispatch is communicated every 4 seconds. However, some entities might need more than 4 seconds to reach desired value due to ramp constraints.
    \item The outcome of real-time deployment does not affect financial rewards collected by entities. Sellers are \textbf{only rewarded for capacity} in the day-ahead market.
\end{itemize}

\subsubsection{Frequency Regulation Problem Formulation}
Based on the outcome of the day-ahead clearing process, the market operator knows for each time step in real-time operations exactly how much capacity is available for deployment by each resource entity. We present a mathematical formulation below for determining how much each entity is dispatched in real-time for any given observed frequency.

Since each hour contains 900 four-second consecutive periods, let $T=900$ be the length of a discrete-time sequential process, which we'll refer to here as an episode, where each step $t\in[0,T-1]$ corresponds to a step in real-time operations. Since the time step is of length 4 seconds and there are 900 steps, the episode is of length 1 hour. Let $t=0$ represent the time where each new hour from the day-ahead market begins. That is, for each real-time episode, the market operator may assume a fixed set of available Reg-Up/Down  capacities for each resource entity, and at the end of time step $T-1$, the episode concludes with a new set of capacities in the episode which follows. Without loss of generality, we focus on only one episode moving forward.

For any episode, let $m$ denote the number of resource entities available for dispatch, i.e. with non-zero cleared capacity from the day-ahead market, corresponding to the same one-hour period, as described in Section \ref{sec:day_ahead}. Let positive scalars $c^\text{up}_1, c^\text{dn}_1, c^\text{up}_2, c^\text{dn}_2, \cdots, c^\text{up}_m, c^\text{dn}_m$ denote the cleared Reg-Up/Down capacities for the $m$ resource entities, and let $d_i[t]$ denote the regulation power dispatched by the market operator at time step $t\in[0,T-1]$ to resource entity $i\in[1,m]$.
Here, $d_i[t]$ represents both Reg-Up and Reg-Down simultaneously; the value is positive during Reg-Down and negative during Reg-Up. This convention is chosen since more power consumption is required when frequency is larger than nominal, i.e. when Reg-Down is needed to pull it down.
One reasonable assumption to make is that there will always be some exogenous fluctuations in the system at each time step that contribute to changes in the frequency, on top of changes in regulation dispatch. Denote all exogenous perturbations from time step $t$ to $t+1$ as $\eta[t]$. Note that this is not a scalar, but an arbitrary object that notationally represents a collection of exogenous forces over an entire 4 second window (i.e. one time step $t$). We first pose the frequency regulation task for each episode as an ideal optimization problem (Eq. (\ref{eq:ideal_optimization})), then we present a set of propositions to simplify the task and yield a practical solution for it.

By convention, boldface refers to vectors. For example $\mathbf{c}^\text{up}\in\mathbb{R}^m$ is the vector of Reg-Up capacities cleared in the day-ahead market for a given episode.

\begin{subequations}
    \label{eq:ideal_optimization}
    \begin{align}
        \begin{split}
            \label{eq:ideal_objective}
            \underset{\{\mathbf{x}[t]\}_{t=0}^{T-1}}{\text{minimize}} \quad& \sum_{t=1}^{T} \max\left(\underline{f}-f[t],0,f[t]-\overline{f}\right)
        \end{split}\\
        \begin{split}
            \text{s.t.} \quad& \forall t \in [0,T-1] ~:
        \end{split}\nonumber\\
        \begin{split}
            \label{eq:ideal_capacity}
            & -\mathbf{c}^\text{up} \leq \mathbf{d}[t+1] \leq \mathbf{c}^\text{dn}
        \end{split}\\
        \begin{split}
            \label{eq:ideal_ramp}
            & \underline{\mathbf{r}} \leq \mathbf{x}[t] \leq \overline{\mathbf{r}}
        \end{split}\\
        \begin{split}
            \label{eq:ideal_x}
            & \mathbf{d}[t+1] = \mathbf{d}[t] + \mathbf{x}[t]
        \end{split}\\
        \begin{split}
            \label{eq:ideal_frequency}
            & f[t+1] = \textbf{PhysicalGrid}\left(f[t],\mathbf{x}[t],\eta[t]\right)
        \end{split}
    \end{align}
\end{subequations}

\vspace{1em}\noindent
where $\mathbf{x}$ denotes incremental changes in Reg-Up/Down dispatch and $\left[\underline{\mathbf{r}},\overline{\mathbf{r}}\right]$ denotes ramp constraints due to physical limitations of resource entities. $f[t]\in\mathbb{R}$ denotes real-time system-wide frequency and $\left[\underline{f},\overline{f}\right]$ is the window of acceptable frequencies. That is, the objective function in Eq. (\ref{eq:ideal_objective}) evaluates to zero if $f[t]\in\left[\underline{f},\overline{f}\right]$ and increases linearly as $f[t]$ escapes the window in either direction.

$\mathbf{x}[t]\in\mathbb{R}^m$ is the decision variable for each step $t$ in the episode, and $f[0],\mathbf{d}[0]$ are exogenous to the optimization problem, yet known. Obtaining an exact solution to Eq. (\ref{eq:ideal_optimization}) is very difficult due to the constraint in Eq. (\ref{eq:ideal_frequency}), which is strictly speaking governed by high-order differential equations, inertia and non-linearities in the system. We would require a rough estimate of the impact of regulation dispatch on the frequency to pragmatically solve this problem. In practice, a heuristic two-stage approach is adopted to addressing this issue. In the first stage, the problem is simplified by decoupling the time steps within each episode so that $\mathbf{x}[t]$ is found separately for each step $t$. It is assumed that all that is needed to find a suitable dispatch is knowledge of system-wide frequency $f[t]$ and operating point $\mathbf{d}[t]$. From this, a required change in dispatch, $\Delta[t]$, is computed proportional to variations in frequency as follows, for some $\beta > 0$:
\vspace{0.5em}
\begin{align}
    \label{eq:dispatch}
    \Delta[t] = \beta\left[ \max\left(0,f[t]-\overline{f}\right) + \min\left(0,f[t]-\underline{f}\right) \right]
\end{align}

Note that if the grid operator desires $\overline{f} = \underline{f} = 60~\text{Hz}$, then the summand in Eq. (\ref{eq:ideal_optimization}) reduces to $\left|f[t] - 60~\text{Hz}\right|$ and the right hand side of Eq. (\ref{eq:dispatch}) reduces to $\beta f[t]$.

Finally, the required net dispatch is distributed across $\mathbf{x}[t]$ while respecting both ramp constraints (Eq. (\ref{eq:ideal_ramp})) and market-cleared capacities for Reg-Up/Down (Eq. (\ref{eq:ideal_capacity})). This restricts $\mathbf{x}$ to the following bounds, at each operating point $\mathbf{d}[t]$:

\begin{subequations}
    \label{eq:x_bounds}
    \begin{align}
        \begin{split}
            \underline{\mathbf{x}}[t] &\gets \max\left(\underline{\mathbf{r}}, -\mathbf{c}^\text{up} - \mathbf{d}[t]\right)
        \end{split}\\
        \begin{split}
            \overline{\mathbf{x}}[t] &\gets \min\left(\overline{\mathbf{r}}, \mathbf{c}^\text{dn} - \mathbf{d}[t]\right)
        \end{split}
    \end{align}
\end{subequations}

At each point in time, the market operator computes required change in dispatch $\Delta[t]$ from Eq. (\ref{eq:dispatch}) along with bounds on incremental dispatch $\left[\underline{\mathbf{x}}[t],\overline{\mathbf{x}}[t]\right]$ from Eq. (\ref{eq:x_bounds}). In the second stage, those values for each time step $t$ are used to solve a simplified problem, posed in Eq. (\ref{eq:pseudo_optimization}) below, to yield $\mathbf{x}[t]$.

\begin{subequations}
    \label{eq:pseudo_optimization}
    \begin{align}
        \begin{split}
            \label{eq:pseudo_objective}
            \underset{\mathbf{x}[t]}{\text{minimize}} & \left|\Delta[t] - \mathbf{1}\cdot\mathbf{x}[t]\right|
        \end{split}\\
        \begin{split}
            \label{eq:pseudo_capacity}
            \text{s.t.} \quad& \underline{\mathbf{x}}[t] \leq \mathbf{x}[t] \leq \overline{\mathbf{x}}[t]
        \end{split}
    \end{align}
\end{subequations}

\vspace{1em}\noindent
where $\mathbf{1}$ denotes a vector with $m$ ones and $\mathbf{1}\cdot\mathbf{y}$ is the sum of entries in some $\mathbf{y}\in\mathbb{R}^m$. The solution to Eq. (\ref{eq:pseudo_optimization}) is proposed in Section \ref{sec:proposed_regulation_dispatch} below.

\subsubsection{Proposed Regulation Dispatch}
\label{sec:proposed_regulation_dispatch}
We propose a pair of possible solutions to Eq. (\ref{eq:pseudo_optimization}) in Propositions \ref{prop:equitable_dispatch} and \ref{prop:sparse_dispatch} below, both of which adopt the principle assumed in Proposition \ref{prop:resource_conservation} below.

\begin{proposition}[Resource conservation]
\label{prop:resource_conservation}
At each point in time, if there's a need for regulation, it is either for increasing or decreasing regulation power, but never for both. Thus, to conserve resources, ensure $\forall t\in[0,T-1]$ that no pair of entries in vector $\mathbf{x}[t]$ are of opposite sign.
\end{proposition}

For any solution, once action $\mathbf{x}[t]$ is determined, the resulting dispatch follows Eq. (\ref{eq:ideal_x}). This is effectively integral droop control, wherein regulation dispatch $\mathbf{d}[t]$ continues to increment/decrement while the frequency is not within the desired range. The only time such form of control is incapable of sufficiently regulating the frequency is when there is simply a shortage of regulation resources. In such cases, the market operator resorts to other forms of ancillary services which are otherwise rarely deployed, such as RRS or NSRS in ERCOT.

There are either one or infinitely many solutions to Eq. (\ref{eq:pseudo_optimization}). If $\Delta[t] \geq \mathbf{1}\cdot \overline{\mathbf{x}}[t]$, the optimal solution is $\mathbf{x}[t] \gets \overline{\mathbf{x}}[t]$. If $\Delta[t] \leq \mathbf{1}\cdot \underline{\mathbf{x}}[t]$, the optimal solution is $\mathbf{x}[t] \gets \underline{\mathbf{x}}[t]$. Infinitely many solutions exist in all other cases, i.e. when $\mathbf{1}\cdot\underline{\mathbf{x}}[t] \leq \Delta[t] \leq \mathbf{1}\cdot\overline{\mathbf{x}}[t]$. We use the term \textit{sufficient resources} in both Proposition \ref{prop:equitable_dispatch} and \ref{prop:sparse_dispatch} below to indicate such cases.

\begin{proposition}[Equitable dispatch]
\label{prop:equitable_dispatch}
Given sufficient resources, i.e. $\mathbf{1}\cdot\underline{\mathbf{x}}[t] \leq \Delta[t] \leq \mathbf{1}\cdot\overline{\mathbf{x}}[t]$, set $\mathbf{x}[t]$ as follows:
\begin{subequations}
    \label{eq:optimal_reg}
    \begin{align}
        \begin{split}
            \label{eq:optimal_reg_up}
            \mathbf{x}[t] &\gets \Delta[t]\cdot\cfrac{\underline{\mathbf{x}}[t]}{\mathbf{1}\cdot\underline{\mathbf{x}}[t]} \quad \text{if} \quad \Delta[t] \leq 0
        \end{split}\\
        \begin{split}
            \label{eq:optimal_reg_dn}
            \mathbf{x}[t] &\gets \Delta[t]\cdot\cfrac{\overline{\mathbf{x}}[t]}{\mathbf{1}\cdot\overline{\mathbf{x}}[t]} \quad \text{if} \quad \Delta[t] > 0
        \end{split}
    \end{align}
\end{subequations}
This ensures that required changes in dispatch are shared in equal proportion among participants, hence the term ``equitable dispatch".
\end{proposition}

\begin{proposition}[Sparse dispatch]
\label{prop:sparse_dispatch}

If $\Delta[t]>0$:
\begin{enumerate}
    \item Sort entries in $\overline{\mathbf{x}}[t]$ in descending order. This yields an ordered list of indices $\mathcal{I}$ which is a permutation of $[1,2,\cdots,m]$ in some new order.
    \item Find the smallest integer $k$ such that \mbox{$\sum_{i=1}^k \overline{x}_{\mathcal{I}_i}[t] \geq \Delta[t]$}.
    \item Then, the solution for each $i^\text{th}$ entry is set as:
    \begin{align*}
        x_{\mathcal{I}_i}[t] \gets
        \begin{cases}
            \overline{x}_{\mathcal{I}_i}[t] & \text{if} \quad i\in[1,k-1] \\
            \displaystyle \Delta[t]-\sum_{j=1}^{k-1} \overline{x}_{\mathcal{I}_j}[t] & \text{if} \quad i=k \\
            0 & \text{otherwise}
        \end{cases}
    \end{align*}
\end{enumerate}
A similar approach is adopted in the case of $\Delta[t]<0$, with appropriate changes in signs and inequalities. The term ``sparse dispatch" refers to the fact that $(m-k)$ resource entities are allocated zero in their entries in $\mathbf{x}[t]$.
\end{proposition}

Proposition \ref{prop:equitable_dispatch} allocates dispatch proportionally over all resource entities. This ensures that all are sharing the burden of deployment, i.e. sacrificing mining for financial reward. However, this also means that many physical resources are used which could lead to more wear-and-tear. To address this issue, we present Proposition \ref{prop:sparse_dispatch} which selects a minimal amount of facilities, those with the fastest response, to collectively meet the required dispatch. While this minimizes the amount of resources used, it has a side effect of punishing larger and/or more nimble facilities, since the market only pays for capacity in the day-ahead market, and not for real-time deployment. This could yield a long term negative effect of less participation by such facilities. As a result, we suggest that the former proposition might be more suitable, but we encourage the reader to consider both.

Later in the paper, we demonstrate how adopting such simple approaches can yield satisfactory results for the physical grid. In the next section below, we explore when it is profitable for mining facilities to adopt such approaches.

\vspace{1em}
\section{Mining Facilities Profit Formulation}
\label{sec:profit_formulation}
In this section, we formulate and quantify the economic benefit of mining facilities participating in Reg-Up/Down. In particular, the question we pose in this section is: when is it profitable for mining facilities to participate in the day-ahead Reg-Up/Down market?

\subsection{Mining Rate of Return Pre-Reg-Up/Down}
To properly quantify a mining facility's profit during participation in the day-ahead Reg-Up/Down market, we ought to first quantify its profit prior to any participation in this market for a fair comparison post-participation. In particular, we consider the case of Bitcoin mining, since Bitcoin is the leading cryptocurrency among those that rely on proof-of-work, i.e. on electric power consumption \cite{WhiteHouse_crypto_report}.

Consider a 24-hour period, and let integer $h\in[0,23]$ denote each hour of this period. In Section \ref{sec:day_ahead}, we let $n$ denote the number of QSEs participating in the day-ahead market, $m\leq n$ of which were cleared with non-zero capacity for dispatch. Without loss of generality, consider that all of those QSEs represent mining facilities, since we are only interested in quantifying the profit for such entities. Let $\lambda_i[h]$ denote average electricity price (in \$/MWh) for QSE $i\in[1,n]$ during hour $h$. That is, the cost of energy incurred by entity $i$ at the end of hour $h$ is the amount of energy it consumed during that hour multiplied by $\lambda_i[h]$.

Let $u^\text{BTC}_i[h]$ denote the long-term utility per Bitcoin mined at hour $h$ as valued by QSE $i$. Moving forward, we express utility in terms of U.S. dollars to perform like-for-like comparisons. Therefore, the units of $u^\text{BTC}_i[h]$ are \$/BTC. Note that we do in fact employ subscript $i$ even though Bitcoin prices are the same for everyone. We do this to distinguish QSEs according to how they perceive the Bitcoins they mine from a long-term perspective. For example, one QSE might elect to immediately sell all Bitcoins it mines, while another might choose to hold on to them for an arbitrarily long time. Hence, although the units of $u^\text{BTC}_i[h]$ are \$/BTC, this quantity is not necessarily equal to the currency conversion rate at hour $h$.

Let $e^\text{BTC}_i[h]$ (in MWh/BTC) denote the total amount of energy required to mine one whole Bitcoin in all facilities collectively represented by QSE $i$ for hour $h$. It is safe to assume that over a 24 hour period, this quantity remains fixed for all hours, but might vary across QSEs, as mining equipment might vary at different facilities. In Section \ref{sec:case_study}, we provide realistic estimates for this quantity.

Define the \textit{mining rate of return} as the value per unit energy returned from mining (in \$/MWh). Let $\rho_i[h]$ denote this amount for QSE $i$ and hour $h$, defined as follows:
\vspace{0.5em}
\begin{align}
    \label{eq:mining_return}
    \rho_i[h] := \cfrac{u^\text{BTC}_i[h]}{e^\text{BTC}_i[h]} - \lambda_i[h]
\end{align}

\vspace{0.5em}\noindent
Note that if electricity price $\lambda_i[h]$ exceeds $u^\text{BTC}_i[h]/e^\text{BTC}_i[h]$ for some hour $h$, then the mining facility is better off not mining. Therefore, in the absence of any Reg-Up/Down participation, the effective rate of return is actually $\max(0,\rho_i[h])$, not $\rho_i[h]$, during hour $h$ assuming that QSE $i$ is acting rationally.

\subsection{Profit Post-Reg-Up/Down}
In Section \ref{sec:day_ahead}, the day-ahead market clearing process was introduced. Let $\pi^\text{up}[h]$ and $\pi^\text{dn}[h]$ (in \$/MW) denote the market-cleared Reg-Up and Reg-Down prices respectively for hour $h$. Those prices are the same for all participants, but the cleared capacities vary across QSEs. Let $c^\text{up}_i[h]$ and $c^\text{dn}_i[h]$ denote the market-cleared Reg-Up and Reg-Down capacities respectively for QSE $i$ at hour $h$, resulting in the following reward collected in the day ahead market:
\vspace{0.5em}
\begin{align}
    R^\text{DAM}_i[h] := \pi^\text{up}[h] c^\text{up}_i[h] + \pi^\text{dn}[h] c^\text{dn}_i[h]
\end{align}

\vspace{0.5em}
Let $C_i[h]$ denote the maximum possible capacity that QSE $i$ is able to either mine with or submit offers with at hour $h$. Indeed it is not necessarily constant over time. For example, the QSE might be involved in some separate demand response program that limits it to some arbitrary amount for that hour. Furthermore, the people behind QSE $i$ might not be willing to offer the full amount each hour. Those are some of the factors that may limit $C_i[h]$.

Now let us say that mining rate of return $\rho_i[h]$ is positive. This means that ideally the mining facilities governed by QSE $i$ should mine as much as possible during hour $h$, which happens to be $C_i[h]$. However, since the QSE is cleared $c^\text{dn}_i[h]$ in day-ahead, it can only mine up to $C_i[h] - c^\text{dn}_i[h]$ to ensure that an amount of $c^\text{dn}_i[h]$ is readily available for Reg-Down. In contrast, if $\rho_i[h] < 0$, then mining any amount at all yields a loss, but since the QSE is cleared $c^\text{up}_i[h]$ in day-ahead, it must mine at least $c^\text{up}_i[h]$ to ensure that an amount of $c^\text{up}_i[h]$ is readily available for Reg-Up. In either case, we can express the net rewards collected in real-time due to mining \textbf{before any real-time deployment} of Reg-Up/Down as follows:
\vspace{0.5em}
\begin{align}
    R^\text{base}_i[h] := \rho_i[h]
    \begin{cases}
    	C_i[h] - c^\text{dn}_i[h] &\text{if} \quad \rho_i[h] \geq 0 \\
    	c^\text{up}_i[h] &\text{if} \quad \rho_i[h] < 0
    \end{cases}
\end{align}
\noindent
This amount strictly decreases for any increase in either $c^\text{up}_i[h]$ or $c^\text{dn}_i[h]$, reflecting a risk being taken during any participation.

Finally, we formulate the rewards collected due to deployment of Reg-Up/Down. In Section \ref{sec:real_time}, we formulated real-time dispatch, wherein we detail how real-time frequency determines dispatch. Since we do not know a priori how much dispatch is needed, we treat it as a random variable in this section. In Section \ref{sec:historical_data}, we explore the distribution of this random variable based on real-world deployment data from ERCOT. Let $\epsilon^\text{up}_i[h]\leq 0$ and $\epsilon^\text{dn}_i[h]\geq 0$ denote deployment rates for Reg-Up and Reg-Down, with the former being strictly non-positive to reflect the fact that Reg-Up deployment reduces consumption. This results in the following additional rewards collected in real-time due to deployment:
\vspace{0.5em}
\begin{align}
    R^\text{Reg}_i[h] := \rho_i[h]\left(\epsilon^\text{up}_i[h]c^\text{up}_i[h] + \epsilon^\text{dn}_i[h]c^\text{dn}_i[h]\right)
\end{align}

\vspace{0.5em}\noindent
Let $R_i[h]$ denote the net rewards, or profit, for QSE $i$ over hour $h$. This is defined simply as the sum of rewards collected in day-ahead and real-time, i.e. $R^\text{DAM}_i[h] + R^\text{base}_i[h] + R^\text{Reg}_i[h]$. To distinguish between the impacts of $c^\text{up}_i[h]$ and $c^\text{dn}_i[h]$ on the net reward, we define two auxiliary variables as follows:
\vspace{0.5em}
\begin{subequations}
    \begin{align}
        \begin{split}
            \label{eq:w_up}
            w^\text{up}_i[h] :=& ~ \pi^\text{up}[h] + \rho_i[h]\left(\epsilon^\text{up}_i[h] + \mathds{1}_{\{\rho_i[h]<0\}}\right)
        \end{split}\\
        \begin{split}
            \label{eq:w_dn}
            w^\text{dn}_i[h] :=& ~ \pi^\text{dn}[h] + \rho_i[h]\left(\epsilon^\text{dn}_i[h] - \mathds{1}_{\{\rho_i[h]\geq0\}}\right)
        \end{split}
    \end{align}
\end{subequations}

\vspace{0.5em}
Auxiliary variables $w^\text{up}_i[h]$ and $w^\text{dn}_i[h]$ effectively indicate how \textit{worthwhile} it is for QSE $i$ to participate in Reg-Up/Down during hour $h$. In the next subsection, we address when it is worthwhile to participate, and by how much capacity. Using those variables, we can express net rewards concisely as:
\vspace{0.5em}
\begin{align}
    \label{eq:net_rewards}
    \Rightarrow ~ R_i[h] &= w^\text{up}_i[h] c^\text{up}_i[h] + w^\text{dn}_i[h] c^\text{dn}_i[h] \nonumber\\
                        &+ C_i[h]\max(\rho_i[h],0)
\end{align}

\noindent
The expression in Eq. (\ref{eq:net_rewards}) contains three distinct terms. The first two suggest that for each of Reg-Up or Reg-Down, it is only worthwhile to have been cleared for capacity if the corresponding auxiliary variables are positive. The last term is completely independent of cleared capacities, and is equal to revenue without participation in Reg-Up/Down whatsoever.

\subsection{When to Participate in Reg-Up/Down}
The only time QSE $i$ can make a decision regarding its participation in Reg-Up/Down is once a day in the day-ahead market. Once it submits price-capacity offers, it can no longer control (a) at what prices or capacities it is cleared, or (b) how much it is deployed in real-time. Nonetheless, each QSE $i$ can make an informed decision based on $R_i[h]$ for each hour $h$ since market operators, such as ERCOT, tend to publish historic records of deployment, prices and capacities cleared. ERCOT also publishes year-ahead inelastic Reg-Up/Down demand for each hour. Thus, the decision of whether to participate in the day-ahead market boils down to whether $\expected{w^\text{up}_i[h] c^\text{up}_i[h] + w^\text{dn}_i[h] c^\text{dn}_i[h]}>0$ (i.e. profitable) or not. In Section \ref{sec:historical_data}, we explore real-world hourly data for multiple years to demonstrate when it is profitable.

The amount of potential reward collected is dependent on how much capacity the QSEs are cleared for Reg-Up/Down in any given hour. This can be expressed as the solution to a convex optimization problem as follows. First, for each QSE $i$ and hour $h$, consider the problem below.
\vspace{1em}
\begin{subequations}
    \label{eq:convex}
    \begin{align}
        \begin{split}
            \label{eq:convex_objective}
            \underset{c^\text{up}_i[h],c^\text{dn}_i[h]}{\text{maximize}} \quad& \expected{w^\text{up}_i[h] c^\text{up}_i[h] + w^\text{dn}_i[h] c^\text{dn}_i[h]}
        \end{split}\\
        \begin{split}
            \label{eq:convex_leq_C}
            \text{s.t.} \quad& c^\text{up}_i[h] + c^\text{dn}_i[h] \leq C_i[h]
        \end{split}\\
        \begin{split}
            \label{eq:convex_up_geq_0}
            & c^\text{up}_i[h] \geq 0
        \end{split}\\
        \begin{split}
            \label{eq:convex_dn_geq_0}
            & c^\text{dn}_i[h] \geq 0
        \end{split}
    \end{align}
\end{subequations}
\vspace{0.5em}

\noindent
The constraint in Eq. (\ref{eq:convex_leq_C}) is justified as follows. For a given cleared capacity $\left(c^\text{up}_i[h],c^\text{dn}_i[h]\right)$, the QSE is obliged to operate at least at $c^\text{up}_i[h]$ and at most at $C_i[h] - c^\text{dn}_i[h]$ to guarantee delivery of maximum capacity if needed. Thus, $C_i[h] - c^\text{dn}_i[h] \geq c^\text{up}_i[h]$ is required, which is equivalent to Eq. (\ref{eq:convex_leq_C}). If QSE $i$ can determine an estimate for $\expected{w^\text{up}_i[h]}$ and $\expected{w^\text{dn}_i[h]}$ for each hour $h$, as illustrated in Section \ref{sec:case_study}, the objective in Eq. (\ref{eq:convex_objective}) can be approximated as $\expected{w^\text{up}_i[h]} c^\text{up}_i[h] + \expected{w^\text{dn}_i[h]} c^\text{dn}_i[h]$. Finally, this reduces Eq. (\ref{eq:convex}) into a linear program, whose solution can be shown to be the following:
\vspace{0.5em}
\begin{align}
    \label{eq:optimal_c}
    \left(c^\text{up}_i[h],c^\text{dn}_i[h]\right)^* \gets
    \begin{cases}
        \left(C_i[h], 0\right) \quad \text{if} & \expected{w^\text{up}_i[h]} \geq \expected{w^\text{dn}_i[h]} \\
                                                                         & \text{and} ~ \expected{w^\text{up}_i[h]} > 0 \\
        \left(0, C_i[h]\right) \quad \text{if} & \expected{w^\text{dn}_i[h]} > \expected{w^\text{up}_i[h]} \\
                                                                         & \text{and} ~ \expected{w^\text{dn}_i[h]} > 0 \\
        \left(0, 0\right) &\text{otherwise}
    \end{cases}
\end{align}

\vspace{0.5em}\noindent
That is, for a fixed estimate for $\expected{w^\text{up}_i[h]}$ and $\expected{w^\text{up}_i[h]}$, we can conclude two things:
\begin{enumerate}[label=(\alph*)]
    \item It is never optimal to be cleared for both Reg-Up and Reg-Down simultaneously.
    \item The optimal net reward for QSE $i$ at hour $h$ is:
    \vspace{0.5em}
    \begin{align}
        \label{eq:max_reward}
        R^*_i[h] = C_i[h]\max\left(\expected{w^\text{up}_i[h]},\expected{w^\text{dn}_i[h]},0\right)
    \end{align}
\end{enumerate}
\vspace{0.5em}

The intuition behind the three possible solutions in Eq. (\ref{eq:optimal_c}) is the following. $\expected{w^\text{up}_i[h]}$ and $\expected{w^\text{dn}_i[h]}$ are the expected per-MW-capacity profit due to participation in Reg-Up/Down for some hour $h$. Therefore, if neither are positive, then it is simply best not to participate, and if any is positive, then the greater of the two indicates whether to participate in Reg-Up or Reg-Down.
In this paper, we do not further explore a risk-controlled optimization formulation, but in such a case, one could argue that the solution will be a convex combination of the three possible solutions representing a less risky participation in the market. In the next section, we explore a realistic distribution for $w^\text{up}$ and $w^\text{dn}$, and analyze when it is profitable for mining facilities to realistically participate in the Reg-Up/Down day-ahead market.

\vspace{1em}
\section{Viability Analysis: A Texas Case Study}
\label{sec:case_study}
In this section, we present a Texas-based case study which quantifies and illustrates the feasibility of participation by Bitcoin mining facilities in ERCOT's Reg-Up/Down market. First, we rely on historical records of the quantities formulated in Section \ref{sec:profit_formulation} to determine under what circumstances it is ideal for such facilities to participate in the market. Second, we simulate, using a synthetic model of the Texas grid, the impact of the control strategies proposed in Section \ref{sec:real_time} on both the electric grid and the participants involved in the market.

\subsection{Real-World Historical Data}
\label{sec:historical_data}
\begin{figure}[!htbp]
\centering
\includegraphics[width=0.48\textwidth, trim={0 0 0 0}, clip]{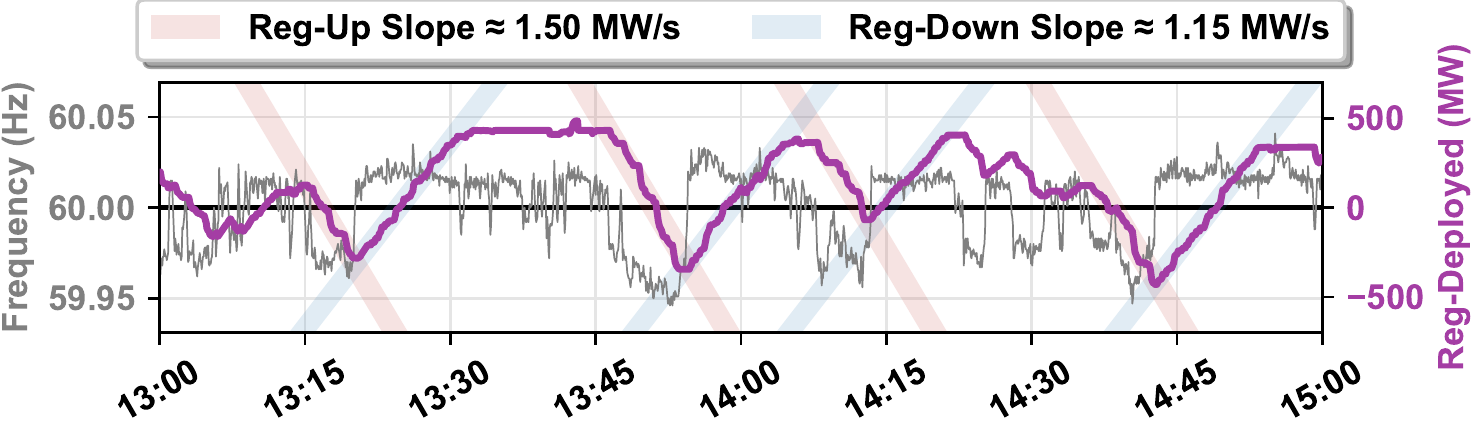}
\caption{Time-synchronized frequency and deployed regulation power, as observed by ERCOT, over a 2-hour window, showing a strong correlation between frequency deviation and change in regulation power.}
\label{fig:freq_reg}
\end{figure}

In Section \ref{sec:real_time}, a formulation for real-time regulation dispatch was presented, wherein a claim was made that simple linear integral control could closely approximate real-world dispatch. In Fig. \ref{fig:freq_reg}, we display a time-synchronized pair of frequency and deployed regulation power, as observed by ERCOT, at a 4-second resolution over a two-hour window sampled from 2022. Here is what we observe from the data:
\begin{enumerate}[label=(\alph*)]
    \item Change in deployed regulation power ($\Delta[t]$ from Section \ref{sec:real_time}) is strongly correlated with frequency deviation from the nominal 60 Hz.
    \item Regulation is deployed even for minuscule frequency deviations (i.e. $\underline{f}$ and $\overline{f}$ from Section \ref{sec:real_time} would be set as very close to 60 Hz).
    \item For extended periods of time, regulation power saturates causing the frequency to remain away from 60 Hz.
    \item There is a visible ramp rate constraint which seems to limit how fast the frequency recovers. The amounts are approximately 1.50 MW/s and 1.15 MW/s for system-wide Reg-Up and Reg-Down respectively.
\end{enumerate}
The mining facilities considered in this paper are a prime candidate at addressing the issue of limited ramping since they are able to ramp up/down consumption for an arbitrary amount of mining machines in parallel in a matter of a few seconds \cite{lancium}. In contrast, as seen in Fig. \ref{fig:freq_reg}, a similar task with existing service providers would take minutes, not seconds, to complete. If we are able to demonstrate that mining facilities can offer capacity with overall profit, then this would address the issue of saturation. We argue this case below.

In Section \ref{sec:profit_formulation}, we concluded that whether or not it is profitable on average for QSE $i$ to participate in Reg-Up/Down during hour $h$ is determined by estimates for $\expected{w^\text{up}_i[h]}$ and $\expected{w^\text{dn}_i[h]}$. For the remainder of this subsection, we quantify such estimates based on historical records of the components that determine these estimates. Such components include \textbf{market cleared prices for capacity}, \textbf{Bitcoin prices}, \textbf{energy to mine a Bitcoin}, \textbf{electricity prices}, and \textbf{deployment rates}. Accordingly we reveal under what circumstances it is profitable to participate in each of Reg-Up and Reg-Down.

\begin{figure}[!htbp]
\centering
\includegraphics[width=0.48\textwidth, trim={0 0 0 0}, clip]{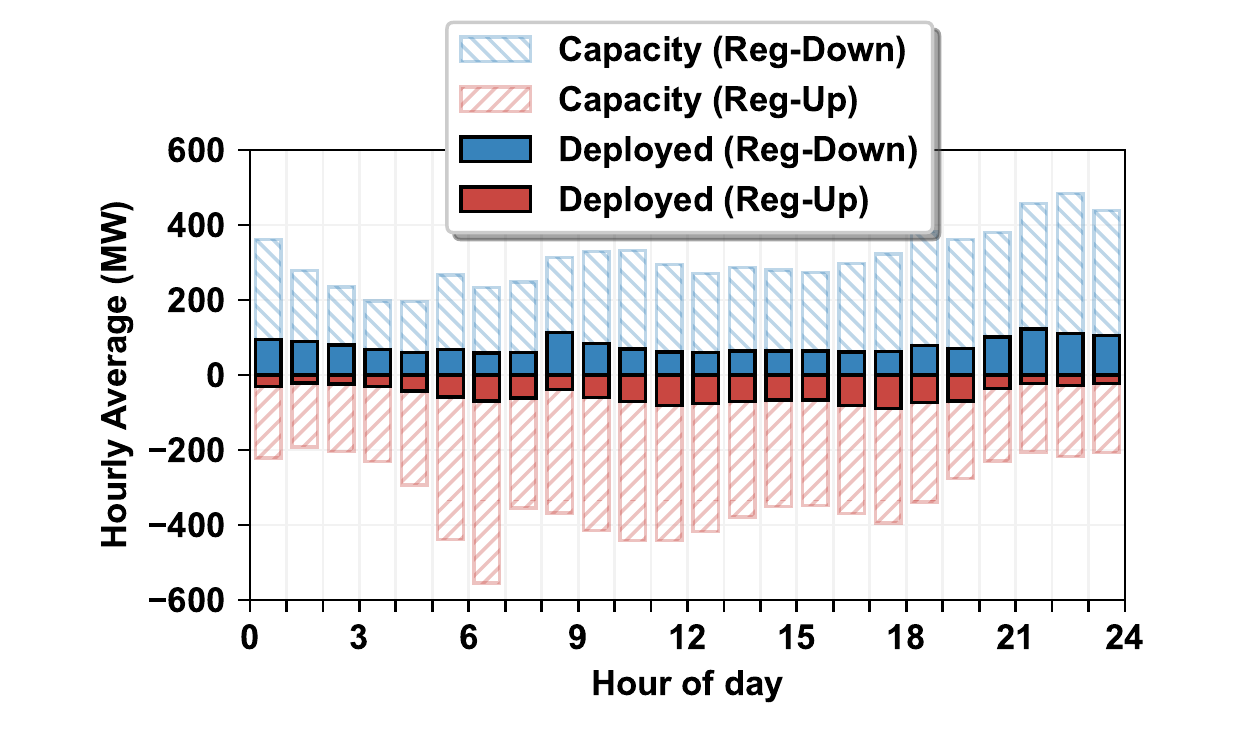}
\caption{Hourly average procured capacity and deployed amounts. Negative values refer to Reg-Up.}
\label{fig:capacity_deployed}
\end{figure}

First, we consider \textbf{deployment rates}, denoted by $\epsilon^\text{up}$ and $\epsilon^\text{dn}$ in the previous section. In Fig. \ref{fig:capacity_deployed}, we show the average capacity procured and average deployed power for each integer hour of the day $h\in[0,23]$. We use negative values for Reg-Up in reference to the fact that Reg-Up corresponds to reduction in consumption, and for visual clarity. From these results, we draw the following conclusions:
\begin{enumerate}[label=(\alph*)]
    \item There are critical hours of the day (e.g. 6-7 AM) where the Reg-Up capacity is much larger than the deployed amount, i.e. $\epsilon^\text{up}$ is very small. Being cleared for Reg-Up capacity during these hours is desirable since it implies that very little opportunity cost is incurred (due to not mining) in exchange for large rewards from the market.
    \item In contrast, as explained in Section \ref{sec:profit_formulation}, a large $\epsilon^\text{dn}$ is desired. Unfortunately, the results reveal $\epsilon^\text{dn}$ is mostly small since Reg-Down deployment (i.e. request to mine more) tends to be much smaller on average than capacity, except for very few hours (e.g. 2-5 AM).
    \item Demand for capacity tends to be quite modest, two orders of magnitude less than demand in the energy market. Consequently, if a few dozen mining facilities were to meet all the required capacity, $c^\text{up}_i[h]$ would be on the order of 10 MW or less on average for each $i$.
\end{enumerate}

\begin{figure}[!htbp]
\centering
\includegraphics[width=0.48\textwidth, trim={0 0 0 0}, clip]{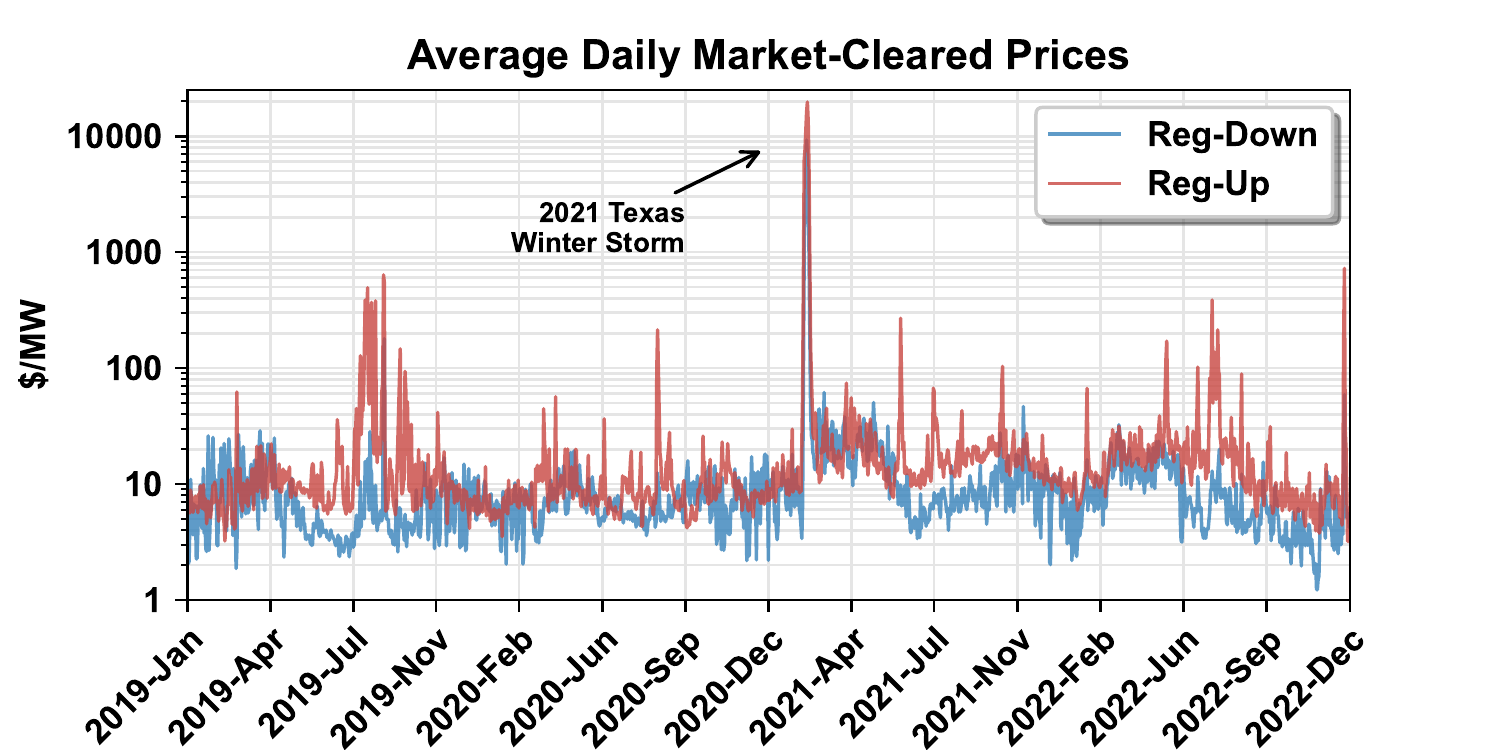}
\includegraphics[width=0.48\textwidth, trim={0 0 0 0}, clip]{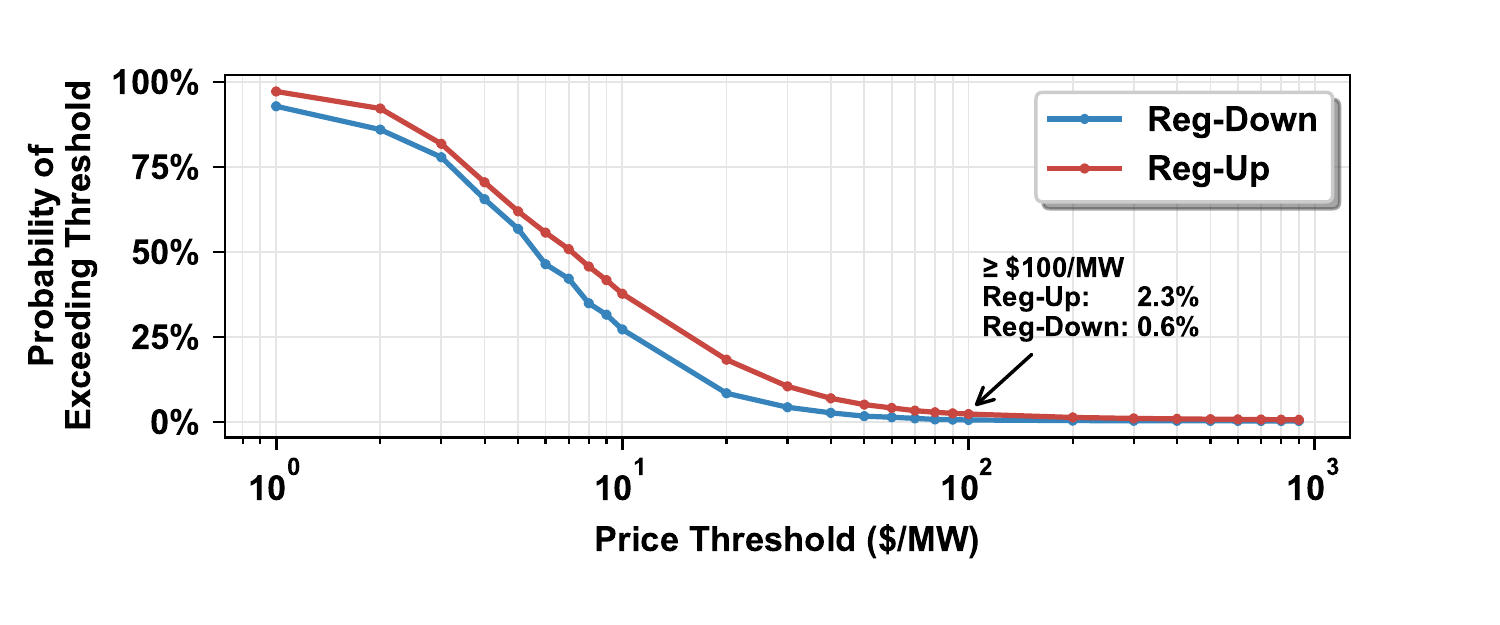}
\caption{Historical market-cleared prices for Reg-Up/Down. Timeline of prices (top), and probability of exceeding different price thresholds (bottom).}
\label{fig:reg_prices}
\end{figure}

Second, we consider \textbf{market cleared prices for capacity}, denoted by $\pi^\text{up}$ and $\pi^\text{dn}$ in the previous section. In Fig. \ref{fig:reg_prices}, we reveal that even in non-extreme cases, the prices, especially for Reg-Up, can reach sufficiently large amounts that it becomes more profitable to forgo mining in exchange for regulation service provision. For example, as shown in the bottom part of the figure, about 2.3\% of the time, Reg-Up prices for capacity exceeded \$100/MW. 
This number by itself holds no meaning; rather, it's strictly about whether it is greater than the opportunity cost due to not mining or not. We illustrate this in the following example.

\begin{figure}[!htbp]
\centering
\includegraphics[width=0.48\textwidth, trim={0 0 0 0}, clip]{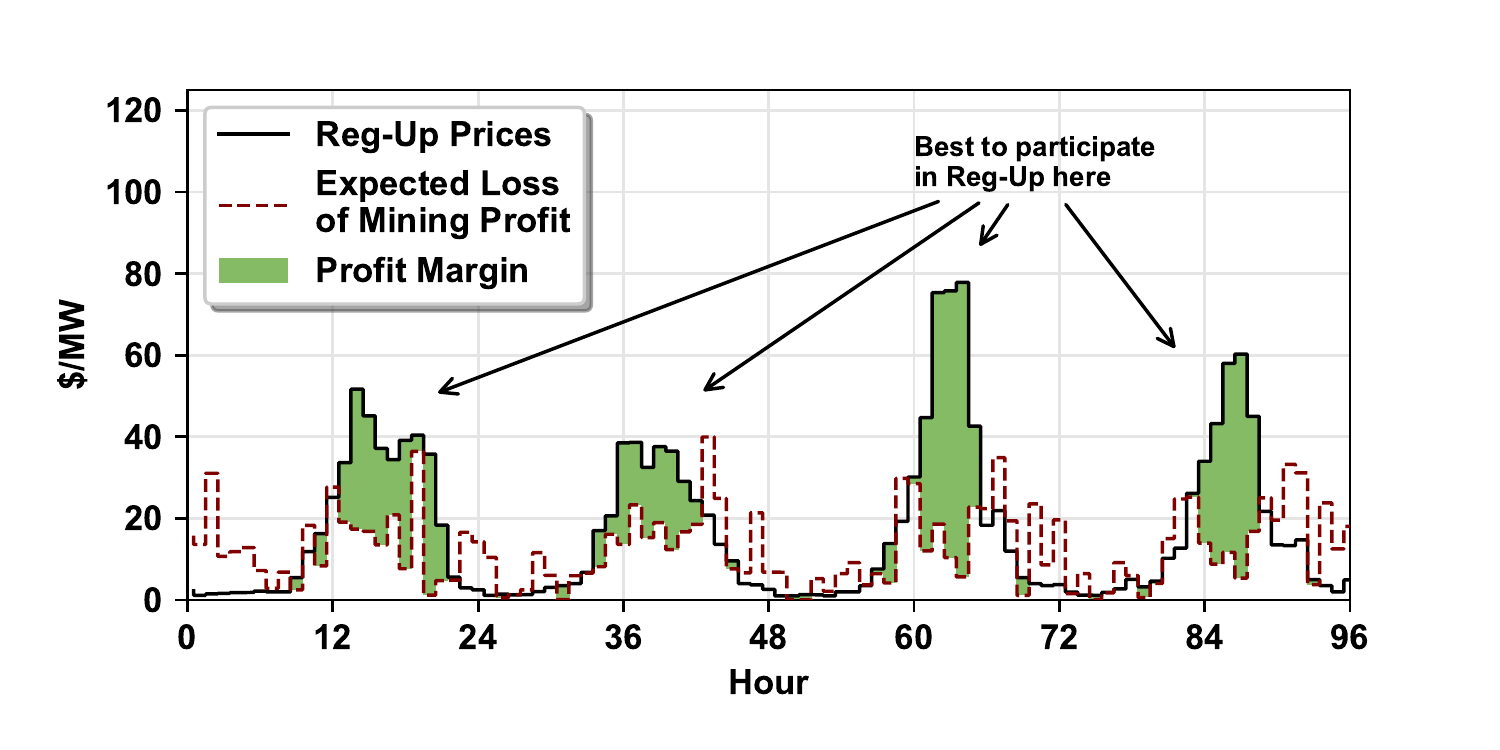}
\caption{An example from June 28 to July 2, 2022 illustrating a potential for profit due to participation in Reg-Up, \textbf{even at low Reg-Up prices}.}
\label{fig:profit_margin_example}
\end{figure}

In Fig. \ref{fig:profit_margin_example}, a 4-day example from the summer of 2022 is shown to illustrate that during certain hours of the day, even when the prices are much less than \$100/MW, it is net profitable to take a risk and to offer Reg-Up capacity. The quantity labelled \textit{Expected Loss of Mining Profit} is effectively the non-price term in Eq. (\ref{eq:w_up}) and is based on deployment rate $\epsilon^\text{up}$. To use the same example, it is evident that deployment rates were sufficiently low during the afternoons such that it was profitable to be cleared for Reg-Up even at low prices. The example shown in Fig. \ref{fig:profit_margin_example} is for less than 1 week of data. Moving forward in this section, we examine the entire historical record spanning four years (Jan 2019 to Dec 2022) to perform a much more exhaustive analysis and determine under what circumstances it is more profitable to have been cleared for capacity than not. Having considered the deployment rates and the market-cleared prices for capacity, what is left is to address is the energy required to mine Bitcoin as well as this cryptocurrency's prices over the years.

Finally, we consider the \textbf{energy required to mine Bitcoin} as well as \textbf{Bitcoin prices} and \textbf{electricity prices}. We rely on assumptions summarized in Table \ref{tab:mining_assumptions} to conclude the following. Consider two state-of-the-art Bitcoin mining machines, released in the summers of 2022 and 2021 respectively: Bitmain Antminer S19 XP and S19J Pro. According to the assumptions made in the table, based on real-world data and blockchain theory, the expected energy required to be awarded one Bitcoin from those two machines amounts to 147 MWh and 202 MWh respectively.

\begin{table}[!htbp]
    \setlength{\tabcolsep}{4pt}
    \centering
    \caption{Assumptions on Bitcoin Mining Capability (Early 2023)}
    \begin{tabular}{rcccc} \toprule
         & \begin{tabular}[c]{@{}l@{}}Quantity\end{tabular}
            \\ \midrule 
        Bitmain Antminer S19 XP (or S19J Pro) & $21.5$ (or $29.5$) $\text{J}/\text{TH}$ \\
        Network Difficulty $D$ (unitless) & $39.35 \times 10^{12}$ \\
        Expected Hashes Per Block & $D/2^{8}~\text{TH}/\text{block}$ \\
        Bitcoin Reward (next halving: Apr. 2024) & $6.25~\text{Bitcoin}/\text{block}$ \\
        \bottomrule
    \end{tabular}
    \label{tab:mining_assumptions}
\end{table}

\begin{figure}[!htbp]
\centering
\includegraphics[width=0.48\textwidth, trim={0 15em 0 10em}, clip]{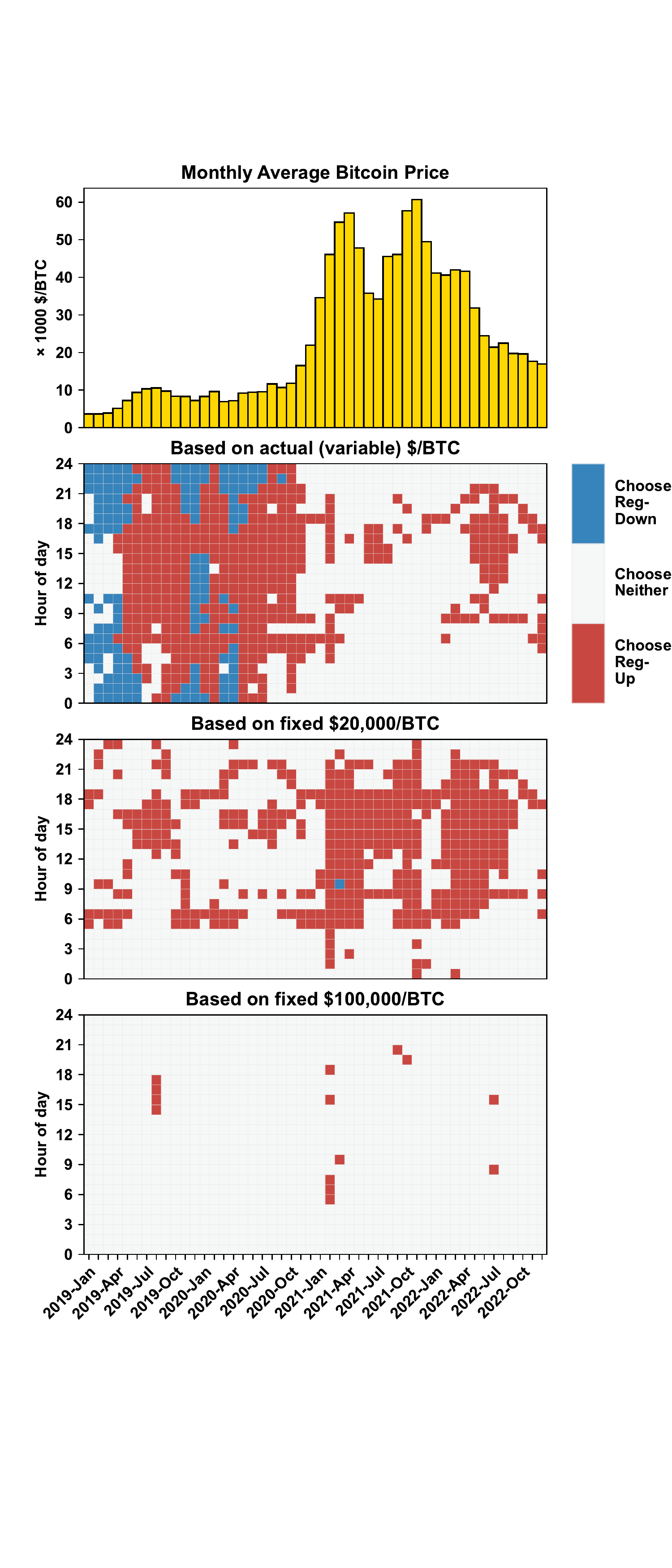}
\caption{Optimal choice (Reg-Up, Down or neither) for each month-hour pair, under different Bitcoin price scenarios, from Jan 2019 to Dec 2022.}
\label{fig:choices_BTC_scenarios}
\end{figure}

For example, at an exchange rate of \$22,050/Bitcoin, the S19 XP's revenue is roughly \$150/MWh. If we consider electricity prices ranging from 5\textcent/kWh to 10\textcent/kWh, the net opportunity cost for 1 hour of cleared capacity ranges from \$50/MW to \$100/MW. This is assuming 100\% deployment for the entire hour. Fortunately, as shown in Fig. \ref{fig:capacity_deployed}, deployment rates are well below 100\%. As a result, the effective opportunity cost reduces to as low as \$10/MW or less. However, if you consider Bitcoin prices greater than \$22,050/Bitcoin, the effective opportunity cost obviously increases. A historical record of average Bitcoin prices is shown in Fig. \ref{fig:choices_BTC_scenarios}.

\begin{figure}[!htbp]
\centering
\includegraphics[width=0.48\textwidth, trim={0 3em 0 0}, clip]{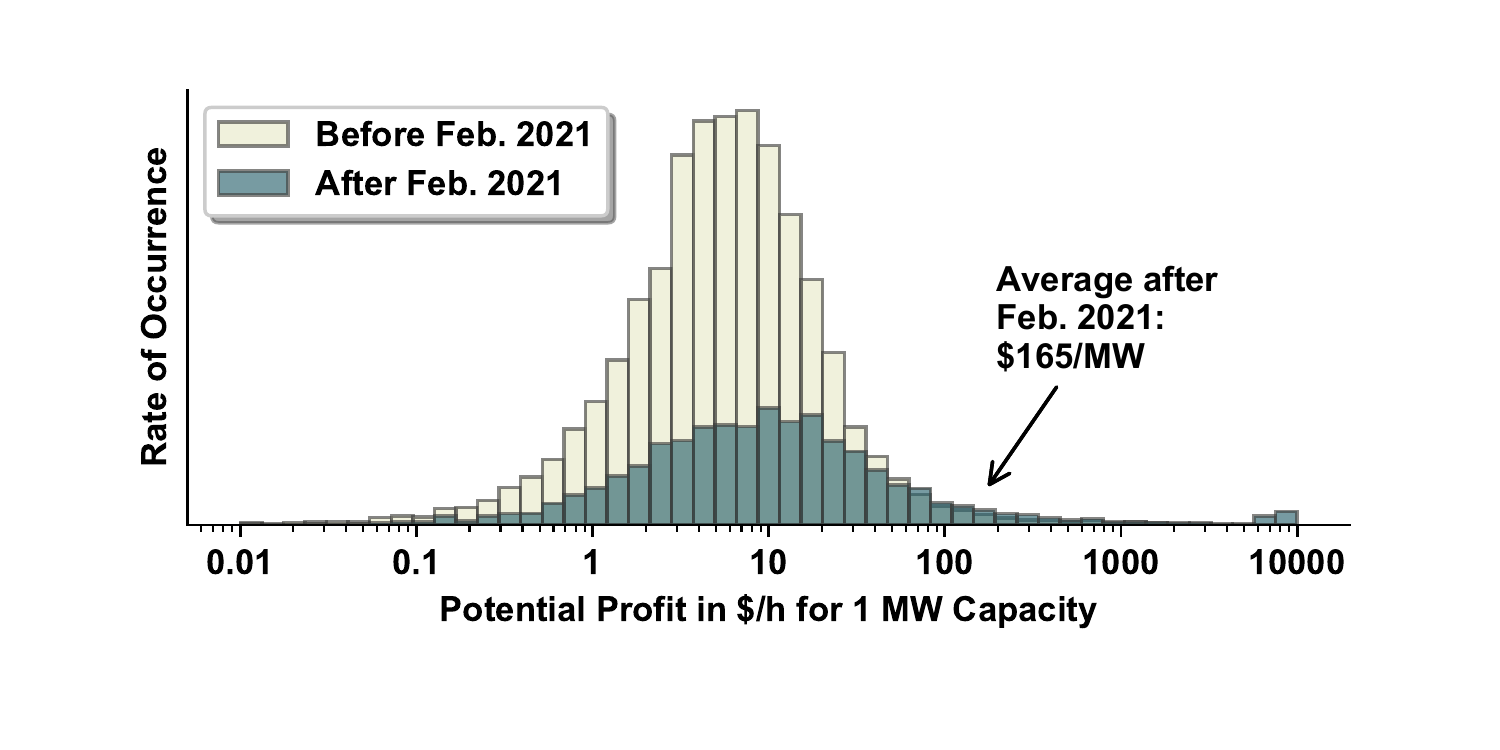}
\caption{Potential profit (Eq. \ref{eq:max_reward}) under proposed strategy (Eq. \ref{eq:optimal_c}). Note that the horizontal scale is logarithmic.}
\label{fig:max_reward}
\end{figure}

As a reminder, to determine whether it is profitable for QSE $i$ to have been cleared for capacity during hour $h$, we need to find when $R^*_i[h] > 0$, as defined in Eq. (\ref{eq:max_reward}). We perform an exhaustive search over the entire 4-year historical record and we reach the conclusion shown in Fig. \ref{fig:max_reward}. We purposely split the data into two: one portion before February 2021 and one after. The reason is twofold: (a) as shown in Fig. \ref{fig:reg_prices}, prices where unusually high during the infamous 2021 Texas winter storm, and (b) as shown in Fig. \ref{fig:choices_BTC_scenarios}, Bitcoin prices where significantly lower before 2021. As a result, when prices were higher, there were much less occurrences where $R^*_i[h] > 0$ after Feb. 2021. Surprisingly, the records reveal that even though there were less occurrences, the average profit was greater during that period, due to increased demand for ancillary services.

Note that the scale in Fig. \ref{fig:max_reward} is logarithmic, which means that the center of mass of the distribution is much farther to the right than where the peak appears to be on the figure. For example, the average is \$165/MW profit after Feb. 2021. Nonetheless, the key takeaway here is that even though profit is much smaller for ancillary services markets than it is for energy markets, it is still worthwhile for mining facilities to consider participation for the following reasons: (a) by providing much faster regulation service and more flexible capacity, they can significantly boost grid resiliency in the face of renewables, which better positions them in the eye of the public, and (b) they do not incur a net loss on average, and the overhead is low to participate since they are already in close communication with local grid operators with minimal communication infrastructure required to set up.

To get a closer look at when it is profitable to participate, the results in Fig. \ref{fig:max_reward} do not suffice. We examine each month-hour pair in the historical record and correspondingly adopt Eq. (\ref{eq:optimal_c}) to choose either to participate in Reg-Up, Reg-Down or neither. Furthermore, we repeat this experiment under three different scenarios for Bitcoin prices: actual historical records, \$20,000/Bitcoin, and \$100,000/Bitcoin. The results are shown in Fig. \ref{fig:choices_BTC_scenarios}, and we draw the following conclusions:
\begin{enumerate}[label=(\alph*)]
    \item There is a clear correlation between increase in Bitcoin prices and increase in choosing not to participate.
    \item The key separating factor between optimally choosing Reg-Up over Reg-Down is time of day.
    \item Even at \$100,000/Bitcoin (never yet witnessed in the historical record), it is still profitable to participate in Reg-Up, but only when the grid is under extreme stress, e.g. during summer heat waves or the 2021 winter storm.
\end{enumerate}

The reason we consider month-hour pairs is based on a protocol followed by ERCOT, wherein, demand for capacity is fixed for each month-hour pair. For example, the demand for Reg-Up capacity during the month of June 2022 at 10 AM is 629 MW for any day of that month, and this was calculated in a year-ahead study \cite{nitika}. Accordingly, we claim that grouping the data in Fig. \ref{fig:choices_BTC_scenarios} by month-hour pairs is appropriate.

\begin{figure}[!htbp]
\centering
\includegraphics[width=0.48\textwidth, trim={0 0 0 0}, clip]{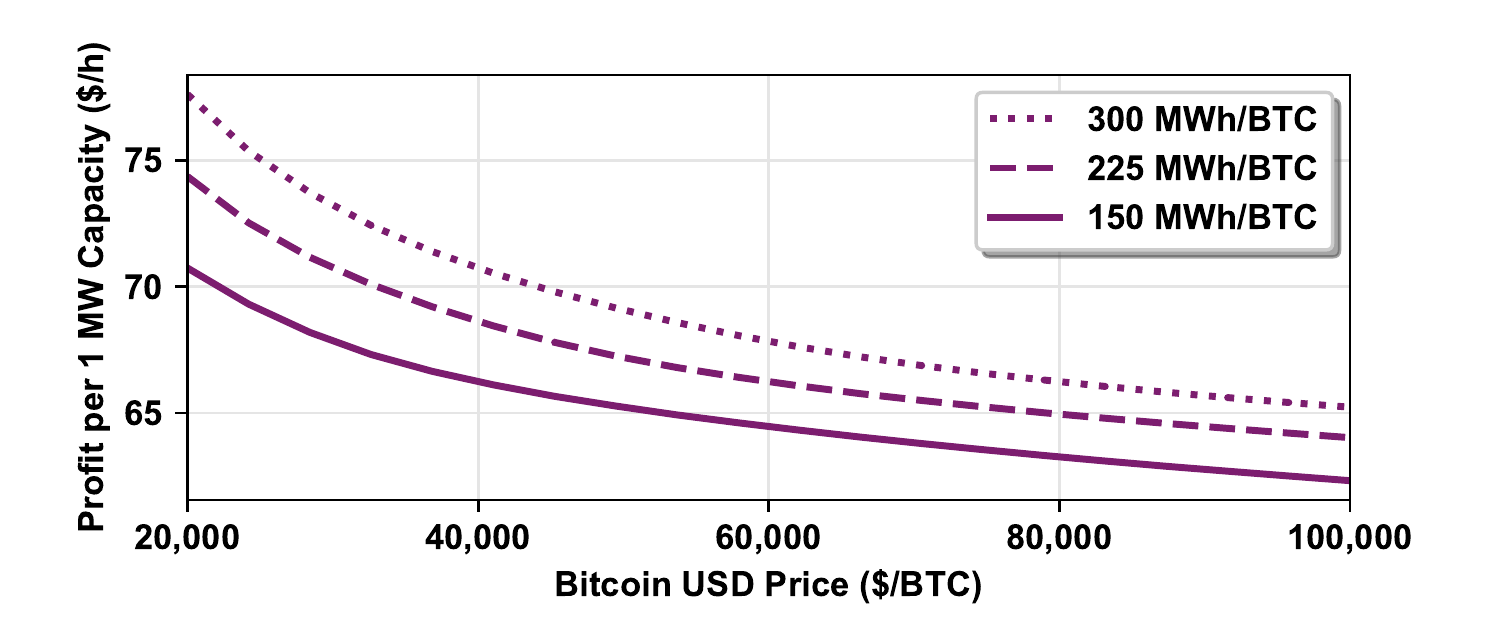}
\caption{Average profits under proposed strategy (Eq. \ref{eq:optimal_c}) for variable Bitcoin prices and multiple scenarios of energy required to mine one Bitcoin.}
\label{fig:profit_per_BTC_and_energy}
\end{figure}

\begin{figure}[!htbp]
\centering
\includegraphics[width=0.48\textwidth, trim={0 0 0 0}, clip]{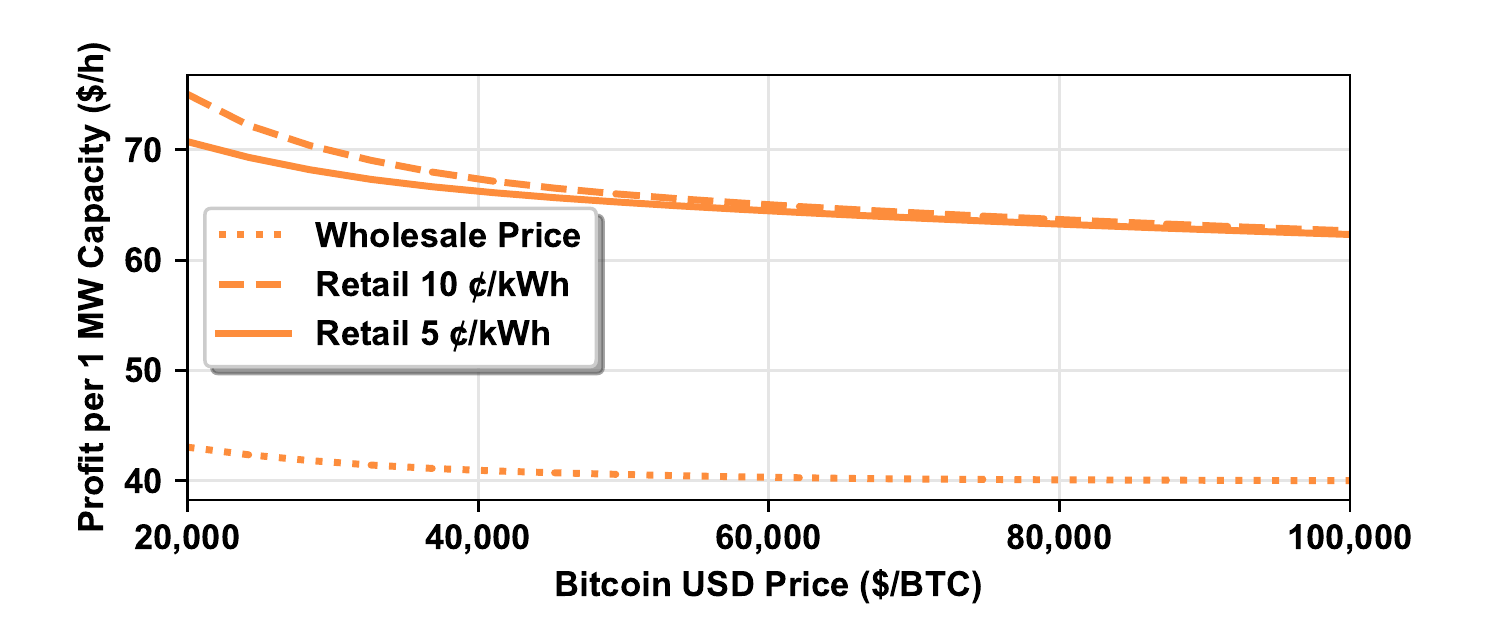}
\caption{Average profits under proposed strategy (Eq. \ref{eq:optimal_c}) for variable Bitcoin prices and multiple scenarios of electricity prices.}
\label{fig:profit_per_BTC_and_electricity_price}
\end{figure}

The result in Fig. \ref{fig:max_reward} reflects a distribution over potential profit assuming that Bitcoin prices followed historical records. In that experiment, we also assumed a fixed 150 MWh/Bitcoin and a retail electricity price of 5\textcent/kWh. To further relax those assumptions, we conduct a series of experiment where we vary all of (a) Bitcoin prices, (b) MWh per Bitcoin and (c) electricity prices, and the results are shown in Fig. \ref{fig:profit_per_BTC_and_energy} and Fig. \ref{fig:profit_per_BTC_and_electricity_price}. In either figure, we sweep over Bitcoin prices from \$20,000/Bitcoin to \$100,000/Bitcoin and plot the average profit per hour under different scenarios. In Fig. \ref{fig:profit_per_BTC_and_energy}, scenarios are defined by energy required to mine one Bitcoin, while in Fig. \ref{fig:profit_per_BTC_and_electricity_price}, scenarios are defined by electricity prices. Based on these results, we conclude that even under extreme situations, it continues to be profitable on average to participate in Reg-Up/Down. For example, for an average cleared Reg-Up capacity of 10 MW, a mining facility can net over \$6M/yr in profit. Note that this profit is calculated relative to the base case of no participation.

\subsection{Texas-Based Simulation}

\begin{figure}[!htbp]
\centering
\includegraphics[width=0.48\textwidth, trim={0 0 3em 0}, clip]{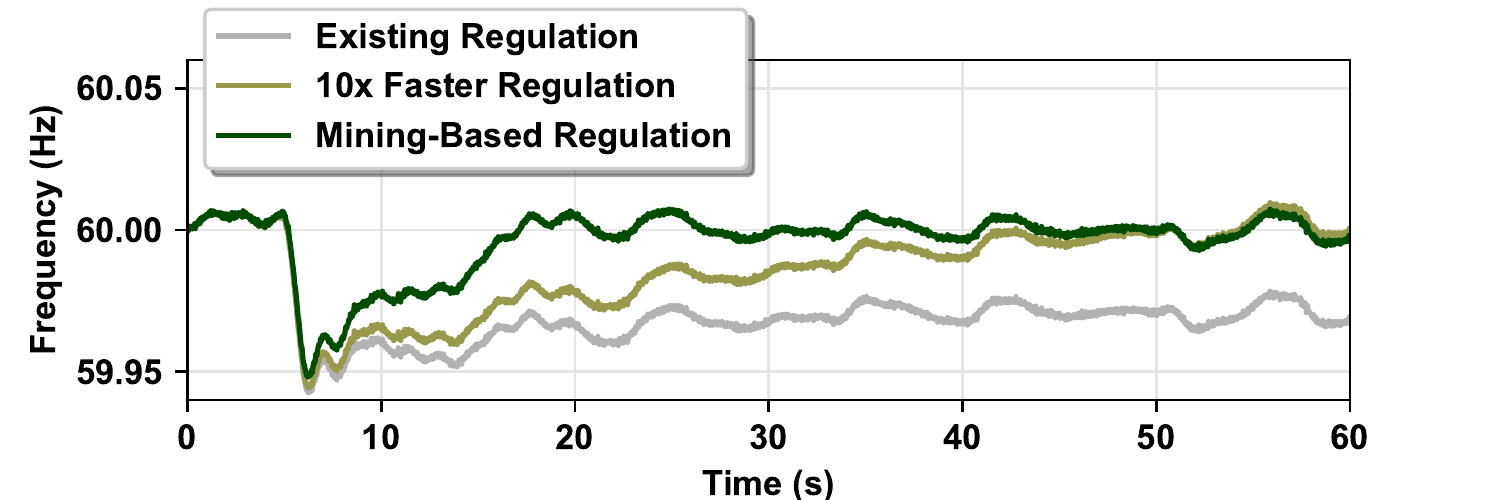}
\caption{Simulated system-wide average frequency in response to loss in generation in a Texas synthetic grid under different scenarios. It takes almost ten times the existing regulation speed to recover the frequency in under a minute. With a total of 500 MW from a collection of mining facilities, the frequency is recovered in much less time.}
\label{fig:frequency_simulation}
\end{figure}

In Section \ref{sec:historical_data}, it was shown that existing ancillary service providers in ERCOT collectively provide Reg-Up support with a speed on the order of 1 MW/s. In this subsection, we present simulation results which reveal that mining facilities could significantly outperform existing providers at the same regulation task. We rely on a synthetic Texas-based 2000-bus transmission network model \cite{Birchfield2017,xu2017creation} to simulate system-wide average frequency at the transient level, and the results are shown in Fig. \ref{fig:frequency_simulation}. In this simulation, there is loss in generation starting at time 5 seconds, and three different strategies for recovering the frequency are compared: (a) using same regulation speed from Fig. \ref{fig:freq_reg} (collectively 1.5 MW/s), (b) using ten times faster regulation, and (c) using a collection of large-scale mining facilities.

For the mining facility, we assume that each individual machine can be ramped down in 10 seconds, which is a conservative estimate according to reference \cite{lancium}.
Considering a collection of large-scale mining facilities which collectively offer 500 MW Reg-Up capacity, this yields an effective ramp rate of 50 MW/s.
The results in Fig. \ref{fig:frequency_simulation} reveal that it takes only a few large-scale mining facilities to outperform even ten times more powerful regulation than what is currently used.

\vspace{1em}
\section{Concluding Remarks}
\label{sec:conclusion}
This paper investigates the potential of utilizing cryptocurrency mining facilities for frequency regulation in electric energy systems. The frequency regulation problem is formulated considering various market participants with different physical constraints, and load dispatch strategies are presented to accommodate these differences. We present market participation strategies that help improve the quality of frequency regulation by exploiting the fast-responding potential of the mining loads while maximizing their operational profit. Our real-world case study and our synthetic transient-level simulations corroborate our theoretical analysis and highlight the advantages of our proposed methods from both economical and physical perspectives. Future work will examine proper ancillary service market designs to incentivize maximum participation of mining facilities in providing grid-level services.



\ifCLASSOPTIONcaptionsoff
  \newpage
\fi


\vspace{1em}
\bibliographystyle{IEEEtran}
\bibliography{references.bib}

\end{document}